\documentclass[prb,twocolumn,amsmath,amssymb,floatfix]{revtex4}
\usepackage{graphicx}
\usepackage{bm}

\newcommand{\be}{\begin{equation}}
\newcommand{\ee}{\end{equation}}
\newcommand{\beq}{\begin{eqnarray}}
\newcommand{\eeq}{\end{eqnarray}}

\begin{document}

\title{Quantum corrections to the dynamics of interacting bosons:
beyond the truncated Wigner approximation.}

\author{Anatoli Polkovnikov}
\email{asp@cmt.harvard.edu}
\homepage{http://athena.physics.harvard.edu/~asp}
\affiliation{Department of Physics, Harvard University, 17 Oxford
St., Cambridge, MA 02138}

\date{\today}

\begin{abstract}
We develop a consistent perturbation theory in quantum
fluctuations around the classical evolution of a system of
interacting bosons. The zero order approximation gives the
classical Gross-Pitaevskii equations. In the next order we recover
the truncated Wigner approximation, where the evolution is still
classical but the initial conditions are distributed according to
the Wigner transform of the initial density matrix. Further
corrections can be characterized as quantum scattering events,
which appear in the form of a nonlinear response of the observable
to an infinitesimal displacement of the field along its classical
evolution. At the end of the paper we give a few numerical
examples to test the formalism.
\end{abstract}

\maketitle

\section{Introduction}

The huge interest to interacting bosons was stimulated by
experimental advances in realization of ultra-cold Bose-Einstein
condensates (BECs)~\cite{Cornell,  Ketterle, Bloch, Bloch1,
Kasevich}. If atoms are a subject to an additional periodic
potential coming from optical standing waves then they form
perfect bosonic crystals with no dislocations, impurities or other
defects. Moreover, the hopping amplitude between the adjacent
cites of this crystal can be varied by orders of magnitude simply
by tuning the intensity of the laser beam~\cite{Bloch,Kasevich}.
Even the sign and the strength of the interaction can be changed
using an external magnetic field~\cite{Cornish, Tiesinga}. So one
can experimentally address various phenomena without worrying
about complications arising from the unwanted degrees of freedom.
For example, in equilibrium, the bosons can undergo a transition
from a superfluid to an insulator as the strength of the optical
potential is increased~\cite{Amico,Elstner, Fisher,Monien,
Jaksh,Sachdev_book}. This transition was directly observed in
Ref.~[\onlinecite{Bloch}]. The simplicity of the system and
relatively slow decoherence suggest a possibility to use the
interacting bosons in quantum computing~\cite{DeMille,Lukin}. A
very appealing idea is to encode quantum information in different
sites of the optical lattice rather than in different internal
atomic states~\cite{Pachos}. In this way the measurement process
and the manipulation of the information seem to be quite
straightforward.

Another huge advantage of the cold atomic systems is that one can
address both theoretically and experimentally dynamic properties
of the interacting atoms. In conventional many-body systems strong
and non-adiabatic perturbations generically lead to fast damping
due to strong coupling to various bath degrees of freedom. So if
one is interested in quantum effects, it is extremely hard to work
far from equilibrium. There are no such limitations, however, in
the atomic systems. Indeed, in order to obtain extremely low
temperatures where the atomic de Broglie wavelength becomes
comparable with inter-particle spacing, it is necessary to isolate
the system to such extent that it can be really thought of as
completely closed. One of the examples of the strongly out of
equilibrium behavior occurs when the sign of interactions changes
from repulsive to attractive, for example using Feshbach
resonance~\cite{Donley, Kanamoto, Kavoulakis}. The important issue
for nonequilibrium problems from the theoretical point of view is
how to take into account quantum fluctuations and go beyond
classical or Gross-Pitaevskii (GP) equations of motion. The latter
can not adequately describe dynamical properties near
instabilities simply because as the fluctuations grow in time, the
initial wavepacket spreads very fast. The obvious generalization
of the GP equations is the Bogoliubov's
approximation~\cite{Leggett_review}. The latter, together with the
perturbative treatment of interactions between quasiparticles,
gives a consistent expansion of the partition function for
equilibrium problems. On the other hand, in nonequilibrium systems
it is not always possible to isolate a unique classical path. In
particular, if the classical motion can not be described by the
linearized version of the GP equations, the whole Bogoliubov's
expansion breaks down. In Ref.~[\onlinecite{ap}] we considered a
specific problem where neither GP nor Bogoliubov's theory would
work. In that paper we studied a condensate, which was
adiabatically driven to the point of the instability, and showed
that because of quantum fluctuations the system evolves into a
macroscopically entangled state. We argued that the correct
dynamics can be reproduced within the truncated Wigner
approximation (TWA)~\cite{Walls, Steel, Lobo, Sinatra, Sinatra1,
Sinatra2}. The idea behind TWA is that the evolution is still
described by the classical GP equations of motion, but the initial
conditions for the classical field are distributed according to
the Wigner transform of the initial density matrix. We showed that
the TWA naturally arises as the first quantum correction to the
classical evolution and it gives asymptotically exact short time
behavior for any bosonic system. The main purpose of the present
investigation is to show how one can go beyond the truncated
Wigner approximation.

There have been some works generalizing GP equations by including
the interaction of the condensate with excited bosons which are
produced during the collapse of the
condensate~\cite{Duine,Stoof1}. Another possible alternative to
the GP and Bogoliubov's approximations is the conventional Keldysh
technique~\cite{Keldysh}. As any diagrammatic expansion, this one
relies on the smallness of interaction both for the classical and
the quantum fields~\cite{Kamenev}. Therefore this technique is not
suitable for strongly interacting systems near the classical
limit. Also it is not suitable for the short-time dynamics, since
the initial conditions are not explicitly written, but rather
absorbed into the quantum propagator in a complicated
manner~\cite{Kamenev}. Quite different class of methods, which
came from quantum optics, are based on the solution of the exact
Fokker-Plank equation for the density matrix which is written in a
coherent state representation~\cite{Gardiner, Walls, Steel}.
Because of the over-completeness of the coherent states, such a
representation is not unique, neither is the resulting
Fokker-Plank equation. For example, in the positive
P-representation, the evolution of the density matrix can be
mapped into the stochastic classical dynamics~\cite{Walls, Steel}.
Within the Wigner approximation, it is easy to reproduce the
correct short-time dynamics and recover TWA, but going further in
time becomes a tricky issue. Recently, there have been developed
an exact stochastic representation of the evolution equations for
the density matrix using the Fock basis, which might be preferable
to the coherent state one~\cite{Carusotto1, Carusotto2}. So far
this theory is restricted to a particular class of two-body
interactions and it is not completely clear (at least to this
author) how to generalize it. Let us also point out that a
semiclassical expansion of the dynamics for a system of
interacting fermions has been discussed in a number of works by
Filinov {\em et. al.}~\cite{Filinov1, Filinov2, Filinov3}. There
the authors suggested a perturbative scheme based on the integral
representation of the evolution equations for the density matrix.
Describing the dynamics in terms of the latter is certainly a
valid concept. However, especially near the classical limit, it
might be preferable to study the corrections to the evolution
equations themselves. And this is exactly what we are going to do
here. Our aim is to develop an expansion similar in spirit to that
given in Refs.~[\onlinecite{Filinov1, Filinov2, Filinov3}].
However, we will give a completely different derivation, which is
suitable for bosonic systems, and will give a simple intuitive
interpretation of the obtained results. The other important
difference between the two approaches is that we will not attempt
to evolve the density matrix itself, all the dynamics will be
ascribed to the observable operators.

Another motivation for studying quantum corrections to the
classical equations of motion is the possibility to take into
account coupling to other external degrees of freedom, which are
usually represented by a thermal bath. Though, as we mentioned
above, the notion of a bath is not particularly useful for the
condensates because of their isolation from the environment, this
coupling certainly provides a strong mechanism of decoherence in
most of condensed matter systems. A standard way to add such a
coupling into the picture is based on the representation of the
bath by a set of harmonic oscillators satisfying certain physical
properties. After integrating out the environment degrees of
freedom one derives Langevin equations, which are essentially
classical equations of motion with extra dissipative and random
force terms~\cite{Caldeira}. Unfortunately, the Langevin equations
can be rigorously derived only for noninteracting particles in a
harmonic potential linearly coupled to the bath, i.e. in the limit
where there are no corrections to the classical equations of
motion due to quantum effects. So incorporating the latter into
the Caldeira-Leggett picture~\cite{Caldeira} is another important
theoretical challenge. Generalization of the derived evolution
equations to open systems will be a subject of our future work.

Before going into the actual calculations let us outline the main
results of this paper. The quantum corrections manifest themselves
in two ways: (i) the initial conditions for the classical
equations of motion are distributed according to the Wigner
transform of the initial density matrix; the classical observable
is the Weyl symbol of the quantum operator, or equivalently its
symmetrized version with fields substituted by c-numbers (see also
discussion preceding~(\ref{last24})), (ii) there are quantum
scattering processes, which are represented as a nonlinear
response of the observable to the infinitesimal transform of the
fields along their classical trajectories. We would like to
emphasize that (i) alone is equivalent to the truncated Wigner
approximation. Amazingly the symmetrized quantum operators and the
Wigner transform appear automatically in the path integral
approach, where initially all the operators are normal-ordered and
there is no obvious reason why the Wigner transform should ever
emerge. For some situations it is sufficient to ignore the quantum
scattering completely and to consider only the evolution along
multiple classical paths.  We would like to emphasize that any
harmonic action (in particular the Bogoliubov's approximation) is
completely described by TWA, i.e. by the classical dynamics with
appropriate initial conditions. Moreover, to recover the
Bogoliubov's approximation it is sufficient to linearize the
classical GP equations around the stationary solution. As we
showed in Ref.~[\onlinecite{ap}], this linearization is not
adequate for problems with unstable dynamics. Quantum scattering
modifies the classical trajectories themselves, but this
modification is conceptually simple. In the first approximation
there is only one scattering event (although distributed over the
whole space and the time interval of the evolution), i.e., one has
to add a small perturbation only once (for a local in time
interaction) during the classical evolution and calculate the
(nonlinear) response of the observable at the end. In the second
order of perturbation theory there are two scattering events, etc.
The whole picture thus resembles the perturbative approach to the
ordinary scattering problem in the Feynman path integral
representation. Although here we mostly concentrate on the
interacting bosons, the results are quite general. They are
applicable to any type of evolution, where the classical equations
of motion arise as a saddle point of the action. Let us also point
out that TWA is equivalent to taking into account {\em all}
classical vertexes in the conventional Keldysh
technique~\cite{Kamenev}. Each quantum scattering event is
equivalent to adding a quantum vertex. The perturbative approach
we develop here should be recovered also in the Fokker-Planck
master equation for the Wigner function~\cite{Steel}, see also
Refs.~[\onlinecite{Filinov1, Filinov2, Filinov3}]. However, to
obtain the nonlinear response of the observable, one must consider
corrections to the distribution function in terms of functional
derivatives with respect to the fields, so that the response is
recovered upon integrating by parts the product of the
distribution function and the observable. While this must be
certainly possible to do, we will see that the functional integral
derivation gives a very simple and elegant way of deriving the
desired equations.

\section{Quantum corrections to classical dynamics}
\label{path}

Let us assume that a system of bosons is described by some
hamiltonian: $\mathcal H(a_j,a^\dagger_j,t)$, which in general
depends on time and is expressed as a polynomial in creation and
annihilation operators $a^\dagger$ and $a$. It does not matter
whether $a$ is a continuous in space or a discrete field sitting
on a lattice. For a single band boson Hubbard model, which we keep
in mind for specific illustrations, the hamiltonian reads:
\be
\mathcal H=\sum_j \Biggl[ -J(a_j^\dagger a_{j+1}+a_{j+1}^\dagger
a_j )+ {U\over 2} a_j^\dagger a_j (a_j^\dagger a_j-1) \Biggr],
\label{mi1}
\ee
where $J$ is the tunneling constant between the neighboring sites
and $U$ is the on-site interaction strength. Both $U$ and $J$ can
explicitly depend on time. We also assume that the initial state
of the system is given by a density matrix $\rho_0$:
\be
\rho_0=\sum_\chi P(\chi)\,|\chi\rangle \langle \chi|,
\ee
where $|\chi\rangle$ represents some basis and $P(\chi)$ is the
probability to be in a particular state (this $P$ would coincide
with the Glauber P-function~\cite{Walls} if $\{|\chi\rangle\}$ are
the coherent states). If initially the system is in the pure
state, the sum contains only a single term. For the Hamiltonian
(\ref{mi1}), the classical limit is obtained as the number of
bosons per site $N$ tends to infinity. So we will develop the
expansion in terms of $1/N$, where the latter plays the role of
the Plank's constant. To see the analogy explicitly let us define
\be
a_j=\sqrt{N}\sqrt{n_j}\mathrm e^{i\phi_j}.\label{com1}
\ee
Then, according to standard quantum-mechanical formulas:
\be
[n_j,\phi_j]={i\over N}.
\label{com}
\ee
We had to scale the factor of $\sqrt{N}$ in (\ref{com1}) to have a
unique limit for $n_j$ at $N\to\infty$. It is easy to recognize in
(\ref{com}) the conventional commutation relation between, say,
the position and the momentum with $1/N$ playing the role of
$\hbar$. The reason why the Plank's constant does not enter
(\ref{com}) explicitly is that the phase $\phi_j$ does not have a
classical analogue, while the phase multiplied by $\hbar$ does
have one; it can be related to the angular momentum in the
rotating systems~\cite{Leggett_review}.

A time dependent expectation value of an arbitrary operator
$\Omega$ is given by:
\be
\Omega(t)=\sum_\chi P(\chi) \langle \chi| T_{K\tau}\, \mathrm e^{i
\int_0^t \mathcal H(\tau)d\tau} \Omega \mathrm e^{-i\int_0^t
\mathcal H(\tau)d\tau} |\chi\rangle
\label{last21}
\ee
Here $T_{K\tau}$ denotes time ordering along the Keldysh contour
going from $\tau=0$ to $\tau=t$ and then back to $\tau=0$. The
exponent of the operator is understood in the usual sense of an
infinite product~\cite{Negele}:
\be
\mathrm e^{i \int_0^t \mathcal
H(\tau)d\tau}=\lim_{Q\to\infty}\prod_{q=0}^Q (1+i\,\Delta\tau\,
\mathcal H(\tau_q)),
\ee
where $\tau_q=t\, q/Q$ and $\Delta\tau=t/Q$. The ordering
$T_{K\,\tau}$ requires that the multipliers corresponding to later
times are placed closer to the operator $\Omega$.

Note that contrary to the derivation of the evolution equations
within the Keldysh technique~\cite{Kamenev}, we ascribe dynamics
to the operator $\Omega$ rather then to the density matrix $\rho$.
In the coherent state basis (\ref{last21}) reads:
\begin{widetext}
\beq
&&\Omega(t)=\int D a_f D a_f^\star Da_b Da_b^\star\, \langle
a_{b\,0}|\rho_0 | a_{f\,0}\rangle\, \mathrm e^{-a_{f\,0}^\star
a_{f\,0}+a_{f\,0}^\star a_{f\,1}+i\mathcal H
(a_{f\,0},a_{f\,1}^\star)\Delta\tau}\!\!\dots \mathrm
e^{-a_{f\,Q}^\star a_{f\,Q}}\nonumber\\
&&~~~\Omega(a_{f\,Q}^\star, a_{b\,Q},t)\mathrm e^{a_{f\,Q}^\star
a_{b\,Q}} \mathrm e^{-a_{b\,Q}^\star a_{b\,Q} +a_{b\,Q}^\star
a_{b\,Q-1}-i\mathcal H(a_{b\,Q}, a_{b\,Q-1}^\star)\Delta\tau}\!\!
\dots \mathrm e^{-a_{b\,0}^\star a_{b\,0}}.
\label{last22}
\eeq
\end{widetext}
Here $\mathcal H(a_{f\!,b},a_{f\!,b}^\star,\tau)$ and
$\Omega(a_{f\!,b},a_{f\!,b}^\star,\tau)$ are the normal ordered
hamiltonian $\mathcal H$ and the observable $\Omega$ with
operators $a$ and $a^\dagger$ substituted by complex numbers
$a_{f\!,b}$ and $a_{f\!,b}^\star$ respectively. The expression
above is intentionally written in the discrete form, since we want
to take special care of the boundary effects. Instead of the
fields $a_f$ and $a_b$ propagating forward and backward in time,
it is convenient to introduce their classical ($\psi$) and quantum
($\eta$) combinations: $a_f=\psi+{\eta/ 2},\;a_b=\psi-{\eta/ 2}$.
The names ``classical'' and ``quantum'' are not accidental since
in the classical evolution all trajectories are uniquely defined
and the backward path should be exactly identical to the forward
one. So $a_q(t)\neq 0$ comes entirely due to quantum fluctuations.
After taking the limit $\Delta\tau\to 0$ in (\ref{last22}) we
derive:
\begin{widetext}
\beq
&&\langle \Omega(t)\rangle =\int D\eta D\eta^\star D\psi
D\psi^\star\, \langle \psi_0-{\eta_0\over 2}|\rho |
\psi_0+{\eta_0\over 2}\rangle\,
\Omega(\psi(t)^\star+{\eta(t)^\star\over 2},\psi(t)-{\eta(t)\over
2})\mathrm e^{-|\psi_0|^2-{1\over 4}|\eta_0|^2-{1\over
2}|\eta(t)|^2}\nonumber\\
&&\mathrm e^{{1\over
2}(\eta_0^\star\psi_0-\eta_0\psi_0^\star)}\mathrm e^{\int_0^t
d\tau\, (\eta^\star(\tau)\mathcal L[\psi,\psi^\star,\tau] -
\eta(\tau)\mathcal L^\star[\psi,\psi^\star,\tau])}\,\mathrm
{Exp}\!\!\left(\!i\!\int_0^t\!\! d\tau \sum_{n\geq
1}\sum_{m=0}^{2n+1} {\partial^{2n+1}\mathcal
H(\psi,\psi^\star,\tau)\over
\partial\psi^m\psi^{\star\,
2n+1-m}}{\eta^m_\tau\eta^{\star\,2n+1-m}_\tau\over
2^{2n}m!(2n\!+\!1\!-\!m)!}\right)\!,
\label{last23}
\eeq
\end{widetext}
where $\mathcal L(\psi,\psi^\star,\tau)$ stands for the classical
(GP) differential operator acting on the field $\psi(t)$.
\be
\mathcal L_j[\psi,\psi^\star,\tau]\equiv {d \psi_j\over d
\tau}+i{\delta\mathcal H(\psi(\tau),\psi^\star(\tau))\over
\delta\psi_j(\tau)}.
\label{add1}
\ee
Note again that the operator $\mathcal L$ as well as the fields
$\psi$ and $\eta$ contain spatial indices which we suppressed in
(\ref{last23}) to simplify notations. The products like
$\eta^\star \mathcal L$ in (\ref{last23}) are understood as the
appropriate  sums: $\sum_j \eta_j^\star \mathcal L_j$. Equating
(\ref{add1}) to zero gives the classical equations of motion. In
particular, if $\mathcal H$ is given by the normal ordered version
of~(\ref{mi1}) with $a_j$ and $a_j^\dagger$ substituted by
$\psi_j$ and $\psi_j^\star$ respectively, then $\mathcal
L_j[\psi,\psi^\star,\tau]=0$ is equivalent to the Gross-Pitaevskii
equation:
\be
i{\partial \psi_j\over \partial
t}=-J(\psi_{j+1}+\psi_{j-1})+U|\psi_j|^2\psi_j.\label{31}
\ee
Let us explain in some detail how we arrive from~(\ref{last22}) to
(\ref{last23}). There are several different contributions to the
exponent in (\ref{last22}), which we call $S$. The first one
($S_1$) comes from the terms, which do not involve the hamiltonian
$\mathcal H$:
\begin{widetext}
\beq
S_1\!=\!\sum_{q=1}^{Q-1} \biggl[a_{f q}^\star(a_{f\,q+1}\!-a_{f
q}) -a_{b q}^\star(a_{b q}\!-a_{b\,q-1})\biggr]\!+a_{f 0}^\star
(a_{f 1}\!-a_{f 0})-a_{b Q}^\star (a_{b Q}\!-a_{b\,Q-1})-a_{b
0}^\star a_{b 0}\!+a_{f Q}^\star (a_{b Q}\!-a_{f Q}).
\label{lasts1}
\eeq
\end{widetext}
Here $q$ denotes the discrete time while the spatial indices are
suppressed. The first sum in the continuum limit transforms into
the integral:
\beq
&&\sum_{q=1}^{Q-1} a_{f\,q}^\star(a_{f\,q+1}-a_{f\,q}) -
a_{b\,q}^\star(a_{b\,q}\!-\!a_{b\,q-1})\to\nonumber\\
&&\to \int_0^t d\tau \left(a_f^\star(\tau){\partial
a_f(\tau)\over\partial\tau}-a_b^\star(\tau){\partial
a_b(\tau)\over\partial\tau}\right),
\label{last_int}
\eeq
which under the substitutions $a_f\to \psi+\eta/2$,
$a_b\to\psi-\eta/2$ and after integrating by parts becomes:
\be
\int_0^t\! d\tau\! \left(\eta^\star(\tau){\partial \psi(\tau)\over
\partial \tau}-\eta(\tau){\partial\psi^\star(\tau)\over \partial\tau}\right)
+\psi^\star(t)\eta(t)-\psi^\star_0\eta_0.
\ee
In the continuum limit the first and the second terms after the
sum in  (\ref{lasts1}) clearly go to zero and the last two read:
\beq
&&a_{fQ}^\star (a_{bQ}-a_{fQ})-a_{b\,0}^\star a_{b\,0}=-|\psi_0|^2-{|\eta_0|^2/ 4}\nonumber\\
&& +{1/ 2}(\psi^\star_0 \eta_0+\eta^\star_0\psi_0)
-\psi^\star(t)\eta(t) - {|\eta(t)|^2/ 2}.
\label{lasts4}
\eeq
Combining equations (\ref{lasts1}) - (\ref{lasts4}) we derive:
\beq
&&S_1=\int_0^t d\tau \left(\eta^\star(\tau){\partial
\psi(\tau)\over \partial
\tau}-\eta(\tau){\partial\psi^\star(\tau)\over
\partial\tau}\right)\nonumber\\
&&-|\psi_0|^2-{|\eta_0|^2\over 4}-{|\eta(t)|^2\over 2}+{1\over
2}(\eta^\star_0\psi_0-\psi^\star_0\eta_0).
\eeq
\vspace*{0.2cm}

\noindent The second contribution to the exponent of
(\ref{last22}) comes from the terms containing the hamiltonian.
However, since all of them are proportional to $\Delta\tau$, the
continuum limit becomes trivial and does not give extra boundary
contributions:
\begin{widetext}
\beq
&&S_2=\int_0^t d\tau (\mathcal
H(a_f(\tau),a_f^\star(\tau),\tau)-\mathcal
H(a_b(\tau),a_b^\star(\tau),\tau))=\int_0^t d\tau\nonumber\\
&&\left(\mathcal H(\psi(\tau)\!+\!{\eta(\tau)\over
2},\psi^\star(\tau)\!+\!{\eta^\star(\tau)\over
2},\tau)\!-\!\mathcal H(\psi(\tau)\!-\!{\eta(\tau)\over
2},\psi^\star(\tau)\!-\!{\eta^\star(\tau)\over
2},\tau)\right).\phantom{XX}
\label{lasts2}
\eeq
And finally the last step in our derivation is the expansion of
the expression above in powers of $\eta$:
\be
\mathcal H(\psi+{\eta\over 2},\psi^\star\!+{\eta^\star\over
2})-\mathcal H(\psi-{\eta\over 2},\psi^\star\! -{\eta^\star\!\over
2}) =\eta{\partial \mathcal H(\psi,\psi^\star)\over \partial
\psi}+\eta^\star{\partial \mathcal H(\psi,\psi^\star)\over
\partial \psi^\star} +\sum_{n\geq 1}\sum_{m=0}^{2n+1}
{\partial^{2n+1}\mathcal H(\psi,\psi^\star)\over
\partial\psi^m\psi^{\star\,
2n\!+\!1\!-m}}{\eta^m\eta^{\star\,2n\!+\!1\!-m}\over
2^{2n}m!\,(2n\!+\!1\!-\!m)!}.
\label{lasts3}
\ee
\end{widetext}
We intentionally separated terms linear in $\eta$ and $\eta^\star$
because they enter the classical equations of motion, while the
higher powers contribute quantum corrections. Combining
(\ref{lasts1}), (\ref{lasts2}), and (\ref{lasts3}) we recover
(\ref{last23}).

Let us discuss how quantum fluctuations enter the classical
dynamics. From equation~(\ref{last23}) we see there are two
boundary and one bulk contributions. Thus, integrating out
$\eta(t)$ results only in the modification of the observable
operator:
\be
\Omega_{cl}=\langle \Omega\left(\psi^\star+{\eta^\star/
2},\psi-{\eta/ 2}\right)\rangle.
\label{last24}
\ee
Here the average is taken over $\eta$ with the measure
$\exp(-|\eta|^2/2)$. Note that the original operator $\Omega$
entering (\ref{last22}) must be written in the normal ordered
form. It is easy to check that integrating out fluctuations
according to (\ref{last24}) is equivalent to rewriting $\Omega$ in
the symmetrized form and substituting $a$ and $a^\dagger$ by
$\psi$ and $\psi^\star$. We also note that $\Omega_{cl}$ coincides
with the Weyl symbol of $\Omega$. Let us give a simple example
illustrating this statement choosing $\Omega$ to be a number
operator:
\be
\Omega=a^\dagger a={1\over 2}(a^\dagger a+a a^\dagger)-{1\over 2}.
\label{new1}
\ee
According to (\ref{last24}) we have:
\be
\Omega_{cl}=\langle \psi^\star+{\eta^\star/ 2},\psi-{\eta/
2}\rangle=\psi^\star\psi-{1\over 2},
\label{new2}
\ee
We see that $\Omega_{cl}$ obtained from the normal-ordered
$\Omega$ using (\ref{last24}) is equivalent to substituting the
symmetrized product of $a$ and $a^\dagger$ by $\psi\psi^\star$ in
(\ref{new1}).

Another boundary contribution originates from the field $\eta_0$
corresponding to the initial time. Because of the coupling to
$\psi_0$, this fluctuation introduces a probability distribution
for the classical initial conditions:
\beq
&&p(\psi_0,\psi_0^\star)=\int d\eta_0^\star d\eta_0 \langle
\psi_0-{\eta_0\over 2}|\rho | \psi_0+{\eta_0\over
2}\rangle\nonumber\\
&&~~~~~~~~~~~~~~~~~~~\mathrm e^{-|\psi_0|^2-{1\over
4}|\eta_0|^2}\,\mathrm e^{{1\over
2}(\eta_0^\star\psi_0-\eta_0\psi_0^\star)},
\label{last26a}
\eeq
Note that $p(\psi_0,\psi_0^\star)$ is nothing but a Wigner
transform of the density matrix in the coherent state
representation~\cite{Gardiner,Walls} and therefore it is not a
positively defined quantity. Sometimes $p(\psi_0,\psi_0^\star)$
has a weird nonlocal behavior and the semiclassical limit is
achieved in somewhat nonintuitive way~\cite{Gardiner}, see also
discussion given in Ref.~[\onlinecite{ap}]. If we ignore
corrections to the classical equations of motion coming from the
higher powers of the quantum field $\eta$ or the last exponent in
(\ref{last23}), then the time dependence of the observable
$\Omega$ will be given by:
\be
\Omega(t)=\int d\psi_0^\star d\psi_0\, p(\psi_0,\psi_0^\star)
\Omega_{cl}(t,\psi_0,\psi_0^\star),
\label{last25}
\ee

\vspace*{0cm} \noindent where $\Omega_{cl}(t,\psi_0,\psi_0^\star)$
is evaluated on the classical field $\psi(t)$ satisfying the
initial condition $\psi(0)=\psi_0$. This expression constitutes
the so called truncated Wigner approximation frequently used in
quantum optics~\cite{Walls}.

\begin{figure}[h]
\includegraphics[width=8.5cm]{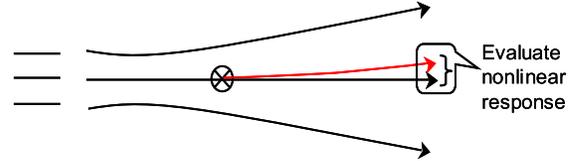}
\caption{Schematic representation of the first quantum correction to
the classical evolution according to (\ref{lastmain}) and
(\ref{lastmain1}). Solid lines represent classical trajectories,
the cross corresponds to a quantum scattering event.}
\label{last:fig0}
\end{figure}

The deviation of the actual trajectories from the classical paths
comes from the bulk quantum fluctuations, which appear in the last
multiplier in (\ref{last23}). Those can be interpreted as quantum
scattering processes. Clearly they are nonzero only if there is an
interaction between the bosons, i.e. $\mathcal H$ contains terms
of the order three or higher in the boson fields $a$ and
$a^\dagger$. For simplicity, let us restrict the following
analysis only to the case of a two particle short range
interaction, which is usually a good approximation for atomic
gases~(see (\ref{mi1})). After the construction is clear, the
generalization on other cases becomes straightforward. So
\be
\mathcal H_{int}={U(t)\over 2}\sum_j a_j^\dagger a_j(a_j^\dagger
a_j-1).
\label{h_int}
\ee
We would like to point out that the unique classical limit at
$N\to\infty$ is obtained when
\be
\lambda(t)\equiv N U(t)
\ee
is kept to be independent of $N$~\cite{psg}. Then the quantum
scattering part of the action reads:
\be
S_q=\int_0^t d\tau {\lambda(\tau)\over 4N}
\left(\psi^\star(\tau)\eta(\tau)+\psi(\tau)\eta^\star(\tau)\right)|\eta(\tau)|^2.
\ee
Because $S_q$ contains the third power (or in general also higher
powers) of the quantum field $\eta$, we can treat it
perturbatively so that:
\begin{widetext}
\beq
&&\Omega(t)=\int d\psi_0^\star d\psi_0 p(\psi_0,\psi_0^\star)
\left[1-i \sum_j\int_0^t d\tau {\lambda(\tau)\over
4N}\left(\psi_j^\star(\tau) {\partial^3\over\partial^2
f^\star_j(\tau)\partial f_j(\tau)} -
\mathrm{c.c.}\right)\right.\nonumber\\
&&~~~~~+{(-i)^2\over 2}\sum_{j,k} \int_0^t\int_0^t d\tau
d\tau^\prime {\lambda(\tau)\lambda(\tau^\prime)\over
16N^2}\left(\psi_j^\star(\tau) {\partial^3\over\partial^2
f^\star_j(\tau)\partial f_j(\tau)} -
\mathrm{c.c.}\right)\nonumber\\
&&~~~~~\times\left.\left.\left(\psi_k^\star(\tau^\prime)
{\partial^3\over\partial^2 f^\star_k(\tau^\prime)\partial
f_k(\tau^\prime)} - \mathrm{c.c.}\right)+\dots\right]
\Omega_{cl}(\psi(t,\{f,f^\star\}),\psi^\star(t,\{f,f^\star\}),t)
\right|_{f,f^\star\equiv 0},
\label{lastmain}
\eeq
\end{widetext}
where $f_j(\tau)$ is an infinitesimal shift of the field
$\psi_j(\tau)$:
\be
\psi_j(\tau)\to\psi_j(\tau)+f_j(\tau),\;
\psi_j^\star(\tau)\to\psi_j^\star(\tau)+f_j^\star(\tau).
\ee
Let us briefly outline the derivation of (\ref{lastmain}).
Expanding (\ref{last23}) in powers of $\eta$ gives the following
integrals:
\be
\int d\eta^\star_\tau d\eta_\tau\, \eta^{\star\,m}_\tau
\eta^n_\tau\mathrm
e^{\eta^\star_\tau(\psi(\tau+\delta\tau)-\psi(\tau)
+iH(\psi(\tau),\psi^\star(\tau),\tau))},
\ee
where $H$ is the classical Hamiltonian of the nonlinear
Schr\"odinger (or GP) equation, not to be confused with $\mathcal
H$:
\be
H(\psi(\tau),\psi^\star(\tau),\tau))={\delta \mathcal
H(\psi(\tau),\psi^\star(\tau),\tau))\over \delta
\psi^\star(\tau)}.
\ee
Next we use the change of variables:
\be
\psi(\tau_q)=\psi(\tau_{q-1})
-iH(\psi(\tau_{q-1}),\psi^\star(\tau_{q-1}),\tau_{q-1})+f(\tau_q)
\label{ttt}
\ee
and a simple identity:
\be
\int x^m \mathrm e^{i\alpha x} dx=2\pi (-i)^m\delta^{(m)}(\alpha).
\ee
Note that the transform from $\psi(\tau_q)$ to $f(\tau_q)$ is
linear and gives no Jacobian for {\em any} interaction. The
classical equations of motion are recovered from~(\ref{ttt}) if
$f\equiv 0$. So nonzero $f$ indeed corresponds to the deviation of
trajectories from the classical ones. Clearly each term in the
expansion in (\ref{lastmain}) gives an extra prefactor of $1/N^2$,
which is the semiclassical parameter for the boson Hubbard model.
To see this, one has to keep in mind that all the fields must be
rescaled as $\psi_j\to \sqrt{N}\psi_j$~\cite{psg}, so that their
expectation values become of the order of $1$ and independent of
$N$. Then each derivative with respect to $f$ or $f^\star$ in
(\ref{lastmain}) would bring an extra factor of $1/\sqrt{N}$. The
interpretation of (\ref{lastmain}) and (\ref{lastmain1}) below
becomes transparent from the figure~\ref{last:fig0}. The solid
lines there denote classical trajectories and the cross represents
a quantum scattering event. The quantum corrections appear as a
nonlinear response to the infinitesimal displacement of the field,
i.e. they reflect the rigidity of the classical motion.
It is straightforward to generalize (\ref{lastmain}) to the
interaction nonlocal in space or time. The only difference is that
the field $\psi$ and its derivatives over $f$ and $f^\star$ would
carry different spatial or time indices. The variables $f$ and
$f^\star$ in (\ref{lastmain}) are treated independently. In
numerical evaluations it is more convenient to use:
\be
{\partial \over\partial f}={1\over 2}{\partial\over \partial \Re
f}-{i\over 2}{\partial\over \partial \Im f},\;
{\partial\over\partial f^\star}={1\over 2}{\partial\over
\partial \Re f}+{i\over 2}{\partial\over \partial \Im f}.
\ee
Then (\ref{lastmain}) becomes:
\begin{widetext}
\beq
&&\Omega(t)=\int d\psi_0^\star d\psi_0 p(\psi_0,\psi_0^\star)
\left[1-{1\over 16} \sum_j\int_0^t d\tau {\lambda(\tau)\over N
}\left({\partial^2\over
\partial\Re f_j^2(\tau)}+{\partial^2\over
\partial\Im f_j^2(\tau)}\right)\right.\nonumber\\
&&\times\left.\left.\left(\Im\psi_j(\tau) {\partial\over\partial
\Re f_j(\tau)} - \Re\psi_j(\tau) {\partial\over\partial \Im
f_j(\tau)}\right)+\dots\right] \Omega_{cl}(\psi(t,\{f\}),t)
\right|_{f\equiv 0}.
\label{lastmain1}
\eeq
\end{widetext}

We would like to make a few comments about
equations~(\ref{lastmain}) and (\ref{lastmain1}). In our specific
example we used the number of bosons per well $N$ as a
dimensionless semiclassical parameter. As we discussed above, each
term in the expansion in those equations brings an extra factor of
$1/N^2$. The absence of $\hbar$ anywhere, may be confusing since
the classical limit certainly corresponds to $\hbar\to 0$.
However, this should not be surprising since here and quite often
in the atomic physics the Planck's constant is either completely
absorbed into energies, which are measured in Hz, or into time. We
already argued above that the ultimate reason why $\hbar$ does not
explicitly appear in our formulas is that the phase does not have
a classical counterpart. If we use conventional observables like
coordinate and momentum or angle and angular momentum,
$\hbar^{-1}$ will appear as a prefactor in the action justifying
the saddle-point or classical approximation. In the same way the
number of bosons per site $N$ appears as a prefactor in the
exponent of (\ref{last23}) after the rescaling
$\psi\to\sqrt{N}\psi$ and $\eta\to\sqrt{N}\eta$. So in general any
expansion in the powers of $\eta$ is in fact the expansion in
powers of $\hbar$. The other important remark is that at small
times the deviation from the classical dynamics due to the quantum
scattering behaves as $O(f(t)/N^2)$, where $f(t)$ is some function
of time, which vanishes at $t\to 0$. This proves that the
truncated Wigner approximation, where the quantum scattering is
completely ignored gives the {\em exact} short-time asymptotical
behavior of the full quantum dynamics. This very remarkable result
is to be contrasted with those obtained within Keldysh technique,
where the short time scales are usually unaccessible.

\section{Numerical examples}

\begin{figure}[ht]
\centering
\includegraphics[width=8.5cm]{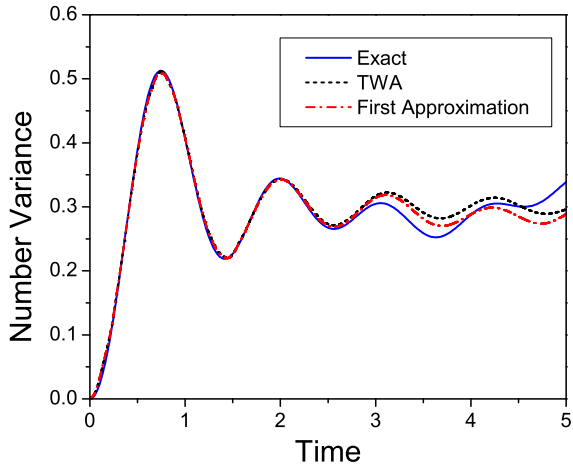}
\caption{Scaled number variance defined in (\ref{examples3}) as a
function of time (measured in the units of inverse tunneling) for
the case of eight bosons per site. TWA includes only fluctuations
at the boundaries of the time domain as described in the text. The
first approximation also includes a single quantum scattering
event.}
\label{last:fig1}
\end{figure}

To illustrate the current approach let us consider some specific
examples. Since the main purpose of this section is not to address
particular problems, but rather numerically test the formalism, we
consider a simple case of two coupled condensates where it is
straightforward to obtain the exact solution and thus to verify
the accuracy of the expansion given in (\ref{lastmain}) and
(\ref{lastmain1}). Formally (\ref{lastmain1}), which we use in
practice, is a multidimensional integral which is best evaluated
using the Monte-Carlo methods. To find the first quantum
correction we apply a small shift to a classical field at a random
moment of time for each set of initial conditions, then follow the
simultaneous evolution of the original and the shifted
trajectories and at the end of the evolution calculate the
response of the observable using finite differences. The first
example we consider is related to the discussion given in our
earlier works~\cite{psg, ap}. Namely, we assume that the two
condensates, described by the hamiltonian (\ref{mi1}) with
$j=L,R$, were initially uncoupled with their wavefunction being a
product of two number states:
\be
|0\rangle=|N\rangle_L |N\rangle_R.
\label{examples1}
\ee
The sub-indices ``L'' and ``R'' correspond to the left and right
sites respectively. It can be shown~\cite{ap} that the Wigner
transform of the density matrix corresponding to the state
(\ref{examples1}) is given by:
\be
p(\psi_0,\psi_0^\star)=2\mathrm e^{-2
(|\psi_{0L}|^2\!+|\psi_{0R}|^2)}L_N(4|\psi_{0 L} |^2)
L_N(4|\psi_{0 R}|^2),
\ee
where $L_N(x)$ stands for the Laguerre's polynomial of the order
$N$. Then suddenly at $t=0$ the tunneling was turned  on and the
following evolution of the system is studied. In
Refs.~[\onlinecite{psg, ap}] we showed that the Gross-Pitaveskii
and the truncated Wigner approximations are very good for
sufficiently short time scales $t\leq t_c\sim N/\lambda=J/U$ (we
remind again that $\lambda=NJ/U$ is the dimensionless measure of
the interaction and the classical limit is achieved at
$N\to\infty$ keeping $\lambda=$const$(N)$) and they break down
completely for longer times $t>t_c$. As an observable it is
convenient to choose a scaled number variance:
\be
\Omega={1\over 4N^2}\left(a_L^\dagger a_L -a_R^\dagger
a_R\right)^2.
\label{examples3}
\ee
The classical counterpart of $\Omega$ can be found either from
direct symmetrization of (\ref{examples3}) or using
(\ref{last24}):
\be
\Omega_{cl}={1\over
4N^2}(\psi_L^\star\psi_L-\psi_R^\star\psi_R)^2-{1\over 8N^2}.
\ee
In equations (\ref{31}) we can always rescale time $t\to t/J$ so
that the dynamics is completely described by a single
dimensionless parameter $\lambda$. For the illustration we will
choose $\lambda=1$ corresponding to the intermediate interaction
strength. In figure~\ref{last:fig1} we plot the resulting
evolution of the number variance for the exact solution, truncated
Wigner approximation and the first quantum correction. Although
account of a single quantum scattering event does not considerably
extend the domain of applicability of the classical description,
it allows to determine the time $t_c$, where the TWA breaks down
without addressing the exact solution. Similarly, the second
correction would show the time scale where the first one breaks
down, etc.
Relatively small extension of $t_c$ for this particular example
should not be surprising. As we showed in Ref.~[\onlinecite{psg}],
the semiclassical description breaks down, because the phase
difference accumulated between the different energy levels becomes
comparable to $\pi$ and the discreteness of the spectrum becomes
crucial. On the other hand in the semiclassical description the
number and the phase are continuously distributed and it becomes
very hard to approximate a discrete sum with essentially chaotic
phases by a continuous integral.

\begin{figure}[ht]
\centering
\includegraphics[width=8.5cm]{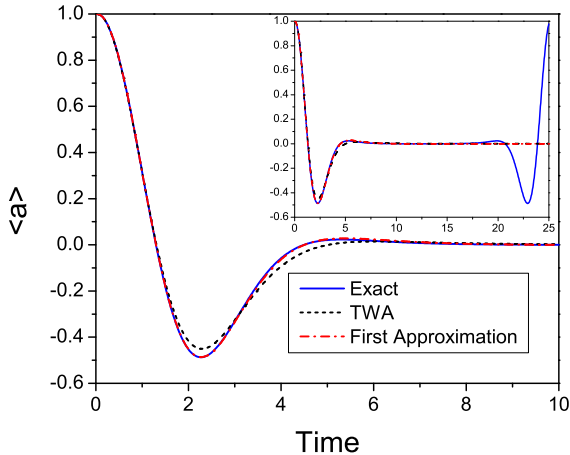}
\caption{Expectation value of the creation operator as a function of time
(measured in units of $1/\lambda$) for the four bosons per well.
The inset shows the same as the main graph, but for a longer time
scale. The semiclassical expansion works very well for the short
time dynamics, but fails to reproduce the restoration of coherence
at a longer time scale. The second quantum correction
(\ref{examples8}) is not shown because it can not be distinguished
by eye from the first one (\ref{examples7}).}
\label{last:fig2}
\end{figure}

Another example where the discreteness of the spectrum is crucial
is quite opposite to that studied above. Assume that initially the
system of bosons was in the noninteracting superfluid state, which
is characterized by the product of coherent states:
\be
|0\rangle=|\sqrt{N}\rangle_{1c} |\sqrt{N}\rangle_{2c}\dots.
\ee
In this example we will not assume that the number of sites is
equal to two, to avoid unnecessary complications coming from the
conservation of the total number of bosons. Then assume that the
tunneling was completely turned off. We will follow the evolution
of the expectation value of $a$ in a particular site. This example
mimics the experiments by M.~Greiner {\em et.
al.}~[\onlinecite{Bloch1}], where collapses and revivals of the
condensate were observed in this way. In the absence of tunneling,
the time evolution in a single site is described by the
hamiltonian:
\be
\mathcal H={\lambda\over N} a^\dagger a(a^\dagger a-1).
\ee
By simple time rescaling we can always choose $\lambda=1$. It is
easy to solve the corresponding Schr\"odinger equation in a number
state so that
\be
\langle a(t)\rangle= \sqrt{N} e^{N (e^{-i t/N}-1) - i t/N}.
\label{examples5}
\ee
It is also straightforward to obtain analytic results within the
truncated Wigner approximation and find the first few quantum
corrections. Thus:
\be
a_{TWA}(t)=\sqrt{N} e^{-\frac{i t}{1 + \frac{it}{2N}}}\frac{1}{\,
    {\left( 1 + \frac{it}{2N} \right) }^2},
\label{examples6}
\ee
\begin{widetext}
\beq
&&a_1(t)=a_{TWA}(t)\left(1 + \frac{t^2}{4N^2} -
    \frac{it^3}{12N^2\left( 1 + \frac{it }{2N}\right)^2 } -
  \frac{it^3}{12N^3\left(1 + \frac{it }{2N}\right)}\right)\label{examples7}\\
&&a_2(t)=a_{TWA}(t)\left(1 + \frac{t^2}{4N^2} -
    \frac{i t^3}{12 N^2\left( 1 + \frac{it }{2N}\right)^2} -
  \frac{i t^3}{12 N^3\left(1 + \frac{it }{2N}\right)}+\frac{t^4}{24N^4}
  - \frac{t^6}{288 N^4{\left( 1 + \frac{it}{2N}\right) }^4}\right.\nonumber\\
&&~~~~~~~~~~~~~~~~~~~~~~~~~\left.- \frac{t^6}{96N^5 {\left( 1 +
\frac{it}{2N}\right) }^3} -
  \frac{7 i t^5}{240 N^4\left( 1 + \frac{it}{2N}\right)^2} -
  \frac{t^6}{192N^6 \left( 1 + \frac{it}{2N}\right)^2} -
  \frac{7 i t^5}{240 N^5\left(1 + \frac{it}{2N}\right)}\right)\label{examples8}.
\eeq
\end{widetext}
From the structure of (\ref{examples6} - \ref{examples8}) it is
clear that the further quantum corrections correspond to the $1/N$
Taylor expansion of (\ref{examples5}) and this indeed can be
explicitly verified. However, both the truncated Wigner
approximation ($a_{TWA}(t)$) and the quantum corrections fail to
reproduce $2\pi N$ periodicity in time of the exact result
(\ref{examples6}). So we can anticipate that the collapse of the
condensate occurring at short time scales can be well described,
while the revivals can not and this is indeed the case. The reason
is that although the function (\ref{examples5}) is analytical in
$N$, one needs to use a number of terms exponentially increasing
with time to correctly reproduce $\langle a(t) \rangle$. On the
physical level we can argue that the restoration of the coherence
comes entirely from the discreteness of the spectrum, so that as
in the first example, the continuous semiclassical expansion does
not work at long time scales. The expectation value of the
creation operator $\langle a\rangle$ as a function of time for the
case of the four bosons per well is plotted in the
figure~\ref{last:fig2}.

\begin{figure}[ht]
\centering
\includegraphics[width=8.5cm]{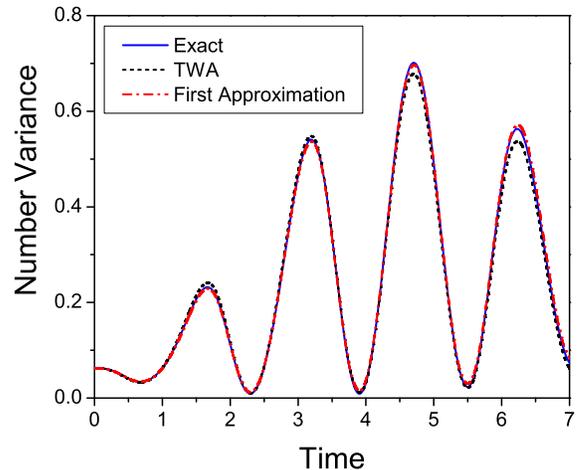}
\caption{Relative number variance as a function of time (measured in units of 1/J)
for a resonant heating of initially noninteracting two coupled
condensates with eight bosons per site. The interaction changes
with time as $\lambda(t)=\sin 4t$. }
\label{last:fig3}
\end{figure}

Now let us consider few more examples, where the suggested
expansion gives considerably better results. Imagine that
initially two non-interacting coupled condensates are a subject to
the oscillating in time interaction:
\be
\lambda(t)=\sin 4t.
\ee
The frequency $\omega=4$ is chosen to be exactly twice that of the
Josephson-like oscillations in the noninteracting system, so that
we can expect a parametric resonance. This process is thus
equivalent to a resonant heating of the condensate. We would like
to point out that the Gross-Pitaveskij approximation completely
fails to describe such a resonance, because the symmetric state is
a classical ground state for any strength of interaction
$\lambda\geq -1$~\cite{ap}. The number conserving Bogoliubov's
approximation also would fail, because near the parametric
resonance we can expect (see figure~\ref{last:fig3}) that a
considerable fraction of particles will be excited. The resulting
dependence of the relative number variance for eight bosons per
well is plotted in figure~\ref{last:fig3}. Obviously, the
truncated Wigner approximation gives a very good description of
the evolution and the first quantum correction does even a better
job so that it hardly deviates from the exact result.

As the last example, we will choose an adiabatic evolution of the
ground state of the two coupled condensates with the increasing
interaction:
\be
\lambda(t)={\tanh \delta t\over 1-\delta t}.
\label{lll}
\ee
Such a form of $\lambda(t)$ is chosen to mimic the tunneling
amplitude $J(t)$ exponentially decreasing with time, with the time
being measured in the units of $J$ (see Ref.~[\onlinecite{ap}]).
Here $\delta$ plays the role of the adiabaticity parameter. If we
again take the number variance as an observable, then it will
simply decrease with time. The question which remains however, is
whether the truncated Wigner approximation will suffice to
correctly describe the evolution or not.
\begin{figure}[ht]
\centering
\includegraphics[width=8.5cm]{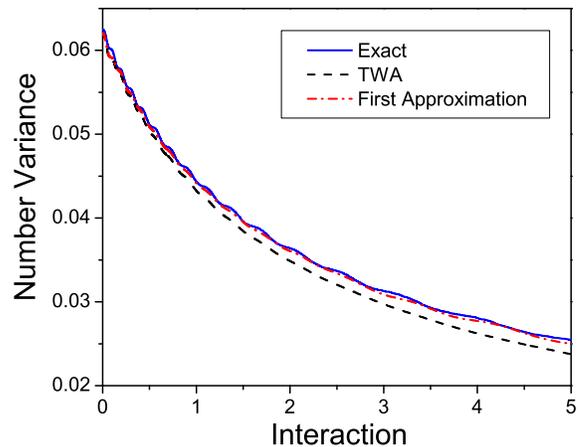}
\caption{Relative number variance as a function of the dimensionless
interaction ($\lambda$) for the interaction increasing with time
according to (\ref{lll}) with $\delta=0.05$. The truncated Wigner
approximation gives a systematic deviation from the exact
solution, while the first quantum correction makes the agreement
with the latter nearly perfect.}
\label{last:fig4}
\end{figure}
In the figure~\ref{last:fig4} we plot the number variance versus
interaction for $\delta=0.05$ and eight bosons per well. It is
obvious that the TWA gives a systematic relative deviation from
the exact solution, which grows in time and the first quantum
correction considerably improves the agreement.

\section{Summary}

Let us now summarize and discuss the derived results. We developed
a time-dependent perturbation theory around the classical
evolution of the system of interacting bosons. We found the two
types of corrections. The first one (which is equivalent to the
truncated Wigner approximation) does not affect the evolution
equations themselves but requires to take an average over an
ensemble of trajectories with initial conditions distributed
according to the Wigner transform of the initial density matrix.
The observable is the classical counterpart of the symmetrized
quantum operator (or its Weyl symbol). Further corrections appear
in the form of quantum scattering processes, which manifest
themselves as a nonlinear response to the infinitesimal change of
the fields along their classical evolution (see (\ref{lastmain})).
We would like to point out again that the widely used Bogoliubov's
approximation, or more generally any expansion of the action up to
the second order in the fields is entirely contained in the TWA.
However, the latter is not limited to the processes, which involve
only stable classical evolution (see also discussion in
Ref.~[\onlinecite{ap}]).

Although we never discussed here the coupling to the thermal bath
(or more generally to the external noise source), it is very
straightforward to incorporate this into our picture. For example,
in a simple representation of the bath as a set of harmonic
oscillators, one obtains a dissipative term, which can be directly
added to the GP equations and a quadratic term in the quantum
fields, which is equivalent to a random force with a Gaussian
distribution~\cite{Caldeira}. In general, there will be other
terms in the action as well, but all of them can be treated
perturbatively in the same way as we explained irrespective of
their origin. So they will result in some kinds of nonlinear
responses to the classical, now stochastic, evolution. Note also,
that the effects of the dissipation or random forces will wash out
all the quantum scattering processes which occur a long time
(longer then the relaxation time) prior to the observation.
Clearly the response to an infinitesimal perturbation will decay
in time if we have such processes. So the theory can be extended
to describe the evolution towards the steady states. One have to
be cautious though in the low temperature limit, because
generically all the relaxation processes are frozen out and the
time scale for the scattering leading to the equilibrium goes to
infinity. This implies that as $T\to 0$ more and more quantum
scattering events have to be considered to correctly describe the
equilibrium. So this theory would give some kind of a high
temperature expansion. If one is interested in non-equilibrium
dynamics, then the limit of the applicability of the perturbative
expansion of the given order will be set either by the time of
evolution or by the scattering time in a bath, whatever is
shorter. Certainly further careful analysis of this approach for
the systems interacting with bath is required and it is a subject
of the future work.

Finally let us spend a few words on the numerical implementation
of the calculations. The solution of the classical equations
themselves is very straightforward. Averaging over the initial
conditions can be easily done with Monte-Carlo methods and the
convergence time either slowly increases with the size of the
system or saturates depending on a particular problem. But in any
case it does not grow exponentially in size as would be the case
for the full quantum solution. The quantum scattering part
requires evaluating a nonlinear response, which might be tricky
for a nonlinear system of differential equations, but this part is
also straightforward and well controlled numerically. We would
like to point out that contrary to exact stochastic schemes, this
method starts directly from the classical equations of motion,
which may be very complicated themselves and shows how to add
quantum corrections step by step. So it certainly must be very
efficient if the quantum fluctuations are relatively weak. Thus
the description of unstable ``cat'' dynamics considered in
Ref.~[\onlinecite{ap}], which is very straightforward in the
present scheme can be hardly efficiently achieved using the
stochastic equations. We believe this method is a competing
alternative both to those based on the conventional diagrammatic
technique, which is not very suitable for strongly interacting
systems, and to the stochastic methods, which can usually deal
with a limited number if degrees of freedom.

The author would like to acknowledge helpful discussions with
I.~Aleiner, I.~Carusotto, A.~Clerk, S.~Girvin, A.~Green,
A.~Melikidze, S.~Sachdev, K.~Sengupta. This research was supported
by US NSF Grant DMR 0098226.

\end{document}